\documentclass[prl,showpacs,preprintnumbers,amsmath,aps,twocolumn]{revtex4}

\def\duzomniejsze{<\kern-.7mm<}
\def\duzowieksze{>\kern-.7mm>}

\def\textbf#1{{\bf #1}}
\def\be{\begin{equation}}
\def\ee{\end{equation}}
\def\bea{\begin{eqnarray}}
\def\eea{\end{eqnarray}}
\def\bse{\begin{subequations}}
\def\ese{\end{subequations}}
\newcommand{\bei}{\begin{itemize}}
\newcommand{\eei}{\end{itemize}}
\newcommand{\bee}{\begin{enumerate}}
\newcommand{\eee}{\end{enumerate}}

\def\hcal{{\cal H}}
\def\tr{{\rm Tr}}

\def\>{\rangle}
\def\<{\langle}

\def\dt#1{{{\kern -.0mm\rm d}}#1\,}

\newtheorem{definition}{Definition}
\newtheorem{proposition}{Proposition}
\begin{document}

\title{Conditional Entanglement}
\author{Dong Yang$^{(1)}$}
\author{Micha\l{} Horodecki$^{(2)}$}
\author{Z. D. Wang$^{(1)}$}

\affiliation{$^{(1)}$Department of Physics and Center of
Theoretical and Computational Physics, The University of
Hong Kong, Pokfulam Road, Hong Kong, China\\
$^{(2)}$Institute of Theoretical Physics and
Astrophysics, University of Gda\'nsk, 80--952 Gda\'nsk,
Poland}

\date{\today}

\begin{abstract}
Based on the ideas of {\it quantum extension} and {\it
quantum conditioning}, we propose a generic approach to
construct a new kind of entanglement measures called {\it
conditional entanglement}. The new measures, built from
the known entanglement measures, are convex,
automatically {\it super-additive}, and even smaller than
the regularized versions of the generating measures. More
importantly, new measures can also be built directly from
measures of correlations, enabling us to introduce an
{\it additive} measure and  generalize it to a
multipartite entanglement measure.
\end{abstract}


\maketitle
%---------------------------------------------------------------------------------------
 Entanglement, as a key resource and ingredient in
quantum information and computation as well as
communication, plays a crucial role in quantum
information theory. It is necessary to quantify
entanglement from different standpoints. A number of
entanglement measures have been formulated, and their
properties have been explored extensively (see, e.g.,
Ref.\cite{4H,PV} and references therein). Nevertheless,
little is known on how to systematically introduce new
entanglement measures. It is likely accepted that an
appropriate entanglement measure is necessarily
non-increasing under local operations and classical
communication (LOCC), while this requirement makes the
definition of entanglement measure notoriously difficult
and challenging. So far, most of existing methods to
construct entanglement measures are based on the "convex
roof" \cite{Bennett1} and the concept "distance"
\cite{Vedral}
--- the distance from the entangled state to its closest
separable state. The well-known entanglement of formation
$E_f$ \cite{Bennett1} is established for a mixed state
$\rho_{AB}$ of a bipartite AB-system via the technique of
convex roof.
%$E_f(\rho_{AB})=\min{\sum_ip_iE(\phi^i_{AB})}$, where the
%minimum is over all pure ensembles
%$\{|\phi^i\>_{AB},p_i\}$ satisfying
%$\rho_{AB}=\sum_ip_i(|\phi^i\>\<\phi^i|)_{AB}$, and
%$E(\phi_{AB})=S(tr_A(|\phi\>\<\phi|)_{AB})$ is the
%entanglement measure for a pure state $\phi_{AB}$, with
%$S(\rho)$ as the von Neumann entropy
%$S(\rho)=-tr\rho\log{\rho}$.
On the other hand, the relative entropy of entanglement
$E_r$  was based on a concept of "distance"
\cite{Vedral}, and squashed entanglement $E_{sq}$  was
built from conditional quantum mutual entropy
\cite{Christandl}---a quantum analog to intrinsic
information \cite{Maurer} known from classical
cryptography, as well as the  logarithmic negativity $E_N
$ was suggested \cite{ZyczkowskiHSP-vol,Vidal} on the
basis of the well-known separability criterion ---partial
transposition \cite{Peres}. Among the known measures,
additivity holds for $E_{sq}$ and $E_{N}$ and is
conjectured to hold for $E_f$, but $E_r$ is nonadditive
\cite{VW}. $E_N$ is computable for a generic mixed state,
while it does not reduce to the von Neumann entropy of
subsystem for pure states. $E_r$ can be generalized to a
measure for multipartite states, but still it is
nonadditive. Very recently, $E_{sq}$ was extended to
multipartite cases \cite{MH}.

In this paper, we introduce a generic approach to
construct a kind of entanglement measures, which is
defined in analogy to the conditional entropy \cite{HOW}
and thus referred to as {\it conditional entanglement}.
The key ideas are quantum extension and quantum
conditioning \cite{HOW}. New  entanglement measures can
be built from old ones and the order between them is
known. Of particular importance, conditional entanglement
can be formulated by quantum conditioning of functions
that describe correlations rather than entanglement.
Taking the quantum mutual information as as exemplary
measure of correlations, we show that a new entanglement
measure can be established by quantum conditioning.
 Remarkably,  it is additive and can straightforwardly be generalized to multipartite states
 for two different choices of multipartite mutual information.

{\definition ~} Let $\rho_{AB}$ be a mixed state on a
bipartite Hilbert space ${\cal H}_{A}\otimes {\cal
H}_{B}$. A conditional entanglement of $\rho_{AB}$ is
defined as \be CE(\rho_{AB})=\inf
\{E(\rho_{AA':BB'})-E(\rho_{A':B'})\}, \ee where the
infimum is taken over all extensions of $\rho_{AB}$,
i.e., over all states satisfying the equation
$\tr_{A'B'}\rho_{AA'BB'}=\rho_{AB}$, and $E(\cdot)$ is an
entanglement measure. Note that the above definition is
similar to that  of conditional entropy
$S(A|B)=S(AB)-S(B)$ with $S(\rho)$ as the von Neumann
entropy $S(\rho)=-\tr\rho\log{\rho}$.

To show that conditional entanglement is a good
entanglement measure, we now elaborate that it does
satisfy two essential axioms that an entanglement measure
should obey \cite{4H}.
\\
{\it 1. Entanglement does not increase under local
operations and classical communication (LOCC) i. e.
$E(\Lambda(\rho))\le E(\rho)$, for any LOCC operation
$\Lambda$.} The reason that $CE$ inherits the
monotonicity of $E$ is straightforward, \be
E(\Lambda^{AB}(\rho_{AA':BB'}))-E(\rho_{A':B'}) \le
E(\rho_{AA':BB'})-E(\rho_{A':B'}).\nonumber \ee {\it 2.
Entanglement is not negative and is zero for separable
states.} The inequality $CE(\rho_{AB})\ge 0$ comes from
the fact that any entanglement measure is non-increasing
by tracing subsystems, while the equality $CE=0$ for
separable states lies in that separable extensions can be
found for separable states.

The monotonicity under LOCC implies that entanglement
remains invariant under local unitary transformations.
This comes from the fact local unitary transformations
are reversible LOCC. The convexity of entanglement used
to be considered as a mandatory ingredient of the
mathematical formulation of monotonicity
\cite{4H,Plenio}. Now the convexity is merely a
convenient mathematical property. Also there is a common
agreement that the strong monotonicity---monotonicity
{\it on average} under LOCC is unnecessary but useful
\cite{4H,Plenio}. Many known existing entanglement
measures are convex and satisfy the strong monotonicity.
We will show that $CE$ naturally inherits these
properties.

For convex $E$, convexity of $CE$ can be obtained by
noticing that for any extension states $\rho_{AA'BB'}$
and $\sigma_{AA'BB'}$, a new extension state can be
constructed as $
\tau_{AA'E:BB'}=\lambda\rho_{AA'BB'}\otimes(|0\>\<0|)_{E}
+(1-\lambda)\sigma_{AA'BB'}\otimes(|1\>\<1|)_{E} $, and
therefore \bea
&~&E(\tau_{AA'E:BB'})-E(\tau_{A'E:B'})\nonumber\\
&=&\lambda [E(\rho_{AA':BB'})-E(\rho_{A':B'})]\nonumber\\
&+&(1-\lambda)[E(\sigma_{AA':BB'})-E(\sigma_{A':B'})].
\eea

Now, let us show that  $CE(\cdot)$ satisfies the
monotonicity on average under LOCC if the convex
$E(\cdot)$ does. It is sufficient to prove that $CE$ is
non-increasing under measurement on one party. For any
extension $\rho_{AA'BB'}$, a measurement on party A
reduces the extension state to an ensemble $\{p_k,
\tilde{\rho}_{AA'BB'}^{k}\}$. \bea
&~&E(\rho_{AA'BB'})-E(\rho_{A'B'})\nonumber\\
&\ge&\sum_{k}p_kE(\tilde{\rho}_{AA'BB'}^{k})-E(\rho_{A'B'})\nonumber\\
&=&\sum_{k}p_kE(\tilde{\rho}_{AA'BB'}^{k})-\sum_{k}p_kE(\tilde{\rho}_{A'B'}^{k})\nonumber\\
&+&\sum_{k}p_kE(\tilde{\rho}_{A'B'}^{k})-E(\rho_{A'B'})\nonumber\\
&\ge&\sum_{k}p_k[E(\tilde{\rho}_{AA'BB'}^{k})-E(\tilde{\rho}_{A'B'}^{k})].
\eea The first inequality comes from the fact that $E$ is
non-increasing on average under local measurement, while
the second one is due to the convexity of $E$. As a
result,  we have $CE(\rho_{AB})\ge \sum_k p_k
CE(\tilde{\rho}_{AB}^k)$.

Remarkably, while most of the known entanglement measures
are sub-additive,  $CE$ is super-additive.

{\proposition ~} $CE(\rho\otimes\sigma)\ge
CE(\rho)+CE(\sigma).$

{\it Proof ~} For any extension state
$\tau_{A_1A_2A':B_1B_2B'}$ of
$\rho_{A_1B_1}\otimes\sigma_{A_2B_2}$, \bea
&~&E(\tau_{A_1A_2A':B_1B_2B'})-E(\tau_{A':B'})\nonumber \\
&=&E(\tau_{A_1A_2A':B_1B_2B'})-E(\tau_{A_2A':B_2B'})\nonumber \\
&+&E(\tau_{A_2A':B_2B'})-E(\tau_{A':B'})\nonumber \\
&\ge&CE(\rho)+CE(\sigma). \eea

Some entanglement measures are upper bounds for
distillable entanglement. Their so-called {\it
regularizations} provide  stronger bounds.  Here $CE$ is
even smaller than the regularized entanglement measure:
\be CE(\rho)\le E^{\infty}(\rho) \quad \text{for all
states } \rho, \label{eq:reg} \ee where
$E^{\infty}(\rho)=\lim_{n\to\infty}E(\rho^{\otimes n})/n$
is the regularized version of the generating entanglement
measure $E$. Indeed, it is explicit that $CE(\rho)\le
E(\rho\otimes |00\>\<00|)-E(|00\>\<00|)=E(\rho).$ From
the super-additivity of $CE$, we know $ nCE(\rho)\le
CE(\rho^{\otimes n})\le E(\rho^{\otimes n}), $ which
leads to (\ref{eq:reg}).

One also finds that $E_f(\rho_{AB:CD})-E_f(\rho_{C:D})\ge
G(\rho_{A:B})$, where $G(\rho_{A:B})>0$ iff $\rho_{AB}$
is entangled \cite{Yang}. We then get $G(\rho_{AB})\le
CE_f(\rho_{AB})\le E_c$, where $E_c=E^\infty_f$ is
so-called entanglement cost \cite{Hayden}. Thus for any
entangled state, $CE_f>0$. It is an open question,
whether $CE_r$ is nonzero for entangled states.

Now let us pass to constructing entanglement measures by
conditioning {\it correlation measures} \footnote{A
measure of correlations  is required at least to be
nonincreasing under local operations,  see
\cite{HendersonVedral}.}. Most intriguingly, we
illustrate below that a new additive measure can indeed
be constructed based on quantum conditioning and can be
generalized to multipartite states.

For a function $f$ quantifying correlations we have two
candidates for its conditioned version \bse\bea
C_f^s(\rho_{AB})&=&\inf[f(\rho_{AA':BB'})-f(\rho_{A':B'})],\\
C_f^a(\rho_{AB})&=&\inf[f(\rho_{A:BE})-f(\rho_{A:E})],
\eea\ese where infimum is taken over all extensions
$\rho_{AA'BB'}$  ($\rho_{ABE}$)  $\rho_{AB}$.
$C_f^s(\cdot)$ is the symmetric conditioned version of
$f$ while $C_f^a(\cdot)$ the asymmetric one.

Taking $f$ to be quantum mutual information
$I(X:Y)=S(X)+S(Y)-S(XY)$, we obtain {\it conditional
entanglement of mutual information} given by $C^s_I$. We
add a factor $1/2$ and will denote it by $C_I$.
Explicitly \be C_{I}(\rho_{AB})=\inf{1\over
2}\{I(AA':BB')-I(A':B')\}, \ee where the infimum is taken
over all the extension states $\rho_{AA'BB'}$ of
$\rho_{AB}$. Now we justify that $C_{I}$ is an
appropriate entanglement measure.

1. We prove that $C_I$ satisfies the strong monotonicity.
From a symmetry consideration, it is sufficient to prove
that $C_{I}$ is non-increasing under a measurement on
subsystem A, namely, $C_{I}(\rho_{AB})\ge \sum_k
p_kC_{I}(\tilde{\rho}_{AB}^{k}),$ where
$\tilde{\rho}_{AB}^{k}=A_{k}\rho_{AB}A_{k}^{\dagger}/p_i$,
$p_i=trA_{k}\rho_{AB}A_{k}^{\dagger}$, and
$\sum_kA_{k}^{\dagger}A_{k}=I_A$. Another way to describe
the measurement process is as following. First, one
attaches two ancillary systems $A_0$ and $A_1$ in states
$|0\>_{A_0}$ and $|0\>_{A_1}$ to system $AB$. Secondly, a
unitary operation $U_{AA_0A_1}$ on $AA_0A_1$ is
performed. Thirdly, the system $A_1$ is traced out to get
the state as $
\tilde{\rho}_{A_0AB}=\sum_{k}A_{k}\rho_{AB}A_{k}^{\dagger}\otimes
(|k\>\<k|)_{A_0}. $ Now for any extension state
$\rho_{AA'BB'}$, we get the state after the measurement
on A, $
\tilde{\rho}_{A_0AA'BB'}=\sum_{k}A_{k}\rho_{AA'BB'}A_{k}^{\dagger}\otimes
(|k\>\<k|)_{A_0}
=\sum_{k}p_k\tilde{\rho}_{AA'BB'}^{k}\otimes
(|k\>\<k|)_{A_0}\nonumber.$  Most crucially, we have
 \bse \bea
&&I(\rho_{AA':BB'})-I(\rho_{A':B'})\nonumber\\
&=&I(0_{A_0A_1}\otimes\rho_{AA':BB'})-I(\rho_{A':B'})\\
&=&I(U_{A_0A_1A}(0_{A_0A_1}\otimes\rho_{AA':BB'}))-I(\rho_{A':B'})\label{se:non1}\\
&\ge& I(\tilde{\rho}_{A_0AA':BB'})-I(\tilde{\rho}_{A':B'})\label{se:non2}\\
%&=&\sum_{k}p_kS(\tilde{\rho}_{AA'}^{k})+S(\tilde{\rho}_{BB'})-\sum_{k}p_kS(\tilde{\rho}_{AA'BB'}^{k})-I(\tilde{\rho}_{A':B'})\nonumber\\
&=&\sum_{k}p_k[I(\tilde{\rho}_{AA':BB'}^{k})-I(\tilde{\rho}_{A':B'}^{k})]\nonumber\\
&+&\sum_kp_k I(\tilde{\rho}_{A':B'}^{k})-I(\tilde{\rho}_{A':B'})\nonumber\\
&+&S(\tilde{\rho}_{BB'})-\sum_{k}p_kS(\tilde{\rho}_{BB'}^{k})\nonumber\\
&=&\sum_{k}p_k[I(\tilde{\rho}_{AA':BB'}^{k})-I(\tilde{\rho}_{A':B'}^{k})]\nonumber\\
&+&\chi(BB')+\chi(A'B')-\chi(A')-\chi(B')\nonumber\\
&\ge&\sum_{k}p_k[I(\tilde{\rho}_{AA':BB'}^{k})-I(\tilde{\rho}_{A':B'}^{k})]\label{se:non3}
\eea \ese where $\chi(\rho)=S(\rho)-\sum_kp_kS(\rho^{k})$
is the Holevo quantity of the ensemble
$\{p_k,\rho^{k}\}$. The equality of (\ref{se:non1}) comes
from that quantum mutual information is invariant under
local unitary operation, while the inequalities of
(\ref{se:non2}) and (\ref{se:non3}) stem from,
respectively, the facts that quantum mutual information
and the Holevo quantity are non-increasing by tracing
subsystem. Consequently, we prove that $C_{I}$ is
non-increasing on average under LOCC operation.

2. $C_{I}\ge 0$ comes from the fact that the quantum
mutual information is non-increasing under tracing
subsystems of both sides. For a separable state
$\rho_{AB}$, it can always be decomposed into a separable
form: $\rho_{AB}=\sum_{i,j}p_{ij}\phi_{A}^{i}\otimes
\phi_{B}^{j}$. An extension state may be chosen to be $
\rho_{AA'BB'}=\sum_{i,j}p_{ij}\phi_{A}^{i}\otimes(|i\>\<i|)_{A'}\otimes\phi_{B}^{j}\otimes(|j\>\<j|)_{B'}.
$ It is obvious that $I(AA':BB')=I(A':B')$, and thus
$C_{I}=0$ for separable states.

{\it Continuity.} The conditional entanglement of quantum
mutual information is asymptotically continuous, i.e. if
$|\rho_{AB}-\sigma_{AB}|\leq \epsilon$, then
$|C_I(\rho)-C_I(\sigma)|\leq K \epsilon \log d +
O(\epsilon)$, where $|\cdot|$ is the trace norm for
matrix, $K$ is a constant, $d=\dim{\hcal_{AB}}$, and
$O(\epsilon)$ is any function that depends only on
$\epsilon$ (in particular, it does not depend on
dimension) and satisfies $\lim_{\epsilon\to
0}O(\epsilon)=0$.

The proof of the asymptotic continuity is similar to that
for the squashed entanglement and is presented in the
Appendix.

{\it Convexity.} $C_{I}$ is convex, i.e., $C_{I}(\lambda
\rho+(1-\lambda)\sigma)\le \lambda
C_{I}(\rho)+(1-\lambda)C_{I}(\sigma)$ for $0\le \lambda
\le 1$.

{\it Proof ~}For any extension states $\rho_{AA'BB'}$ and
$\sigma_{AA'BB'}$, we consider the extension state $
\tau_{AA'A''BB'B''}=\lambda
\rho_{AA'BB'}\otimes(|0\>\<0|)_{A''}\otimes(|0\>\<0|)_{B''}
+(1-\lambda)\sigma_{AA'BB'}\otimes(|1\>\<1|)_{A"}
\otimes(|1\>\<1|)_{B''} $, and have $
I(\tau_{AA'A'':BB'B''})-I(\tau_{A'A'':B'B''}) =\lambda
[I(\rho_{AA':BB'})-I(\rho_{A':B'})] +(1-\lambda)
[I(\sigma_{AA':BB'})- I(\sigma_{A':B'})]. $ This implies
$C_{I}$ is convex.

An immediate corollary of convexity is that $C_{I}\le
E_f$ and furthermore $C_{I}\le E_c$ due to the following
additivity.

{\proposition ~} $
C_{I}(\rho_{AB}\otimes\sigma_{CD})=C_{I}(\rho_{AB})+C_{I}(\sigma_{CD}).
$

{\it Proof ~} On the one hand, for any extension states
$\rho_{AA'BB'}$ and $\sigma_{CC'DD'}$,
$\rho_{AA'BB'}\otimes \sigma_{CC'DD'}$ is an extension
state of $\rho_{AB}\otimes\sigma_{CD}$. \bea
&~&I(AA'CC':BB'DD')-I(A'C':B'D')\nonumber\\
&=&I(AA':BB')-I(A':B')\nonumber\\
&+&I(CC':DD')-I(C':D'). \eea So
$C_{I}(\rho_{AB}\otimes\sigma_{CD})\le
C_{I}(\rho_{AB})+C_{I}(\sigma_{CD})$ holds.

On the other hand,  for extension states
$\tau_{ACE':BDF'}$ of $\rho_{AB}\otimes\sigma_{CD}$,
$\tau_{ACE':BDF'}$ is an extension state of $\rho_{AB}$
and $\tau_{CE':DF'}$ is an extension state of
$\sigma_{CD}$. Therefore we have \bea
&~&I(ACE':BDF')-I(E':F')\nonumber\\
&=&I(ACE':BDF')-I(CE':DF')\nonumber\\
&+&I(CE':DF')-I(E':F'). \eea This means that
$C_{I}(\rho_{AB}\otimes\sigma_{CD})\ge
C_{I}(\rho_{AB})+C_{I}(\sigma_{CD})$. So we have finally
the additivity equality.

It is quite remarkable that the  property of additivity
is rather easy to prove for conditional entanglement
while it is extremely tough for other candidates. The
reason lies in that the conditional entanglement is
naturally  supper-additive while others are  usually
sub-additive. Also the proof for the conditional
entanglement shares a similarity with that of squashed
entanglement. As a matter of fact, squashed entanglement
can be constructed in the same spirit: it is based on
asymmetric conditioning of mutual information \be
E_{sq}(\rho_{AB})={1\over 2}\inf\{I(A:BE)-I(A:E)\}\equiv
{1\over 2}C_I^a(\rho_{AB}), \ee where the infimum is
taken all extensions $\rho_{ABE}$ of $\rho_{AB}$. It is
notable that $I(A:BE)-I(A:E)=I(AE:B)-I(E:B)$ is symmetric
w.r.t. systems $AB$ though each term in the formula is
asymmetric w.r.t. both parties. This gives the
possibility to build symmetric entanglement measures by
asymmetric conditioning.

In \cite{MH}, we call the squashed entanglement {\it
q-squashed entanglement} $E_{sq}^{q}$ because the
extension is generic and the system $E$ is required to be
quantum memory. If we restrict $E$ to classical memory,
another proper entanglement measure---{\it c-squashed
entanglement} $E_{sq}^{c}$ can be obtained~\cite{MH}.
Here we show the order relation among these three
measures.

{\proposition ~\label{se:order}} $E_{sq}^{q}\le C_I\le E_{sq}^{c}$. \\
{\it Proof.} $E_{sq}^{q}\le C_I$ comes from the chain
rule for quantum mutual information. \bea
&~&I(AA':BB')-I(A':B')\nonumber\\
&=&I(A':BB')+I(A:BB'|A')-I(A':B')\nonumber\\
&=&I(A':B|B')+I(A:B'|A')+I(A:B|A'B')\nonumber\\
&\ge&I(A:B|A'B'). \eea The proof of $C_I\le E_{sq}^{c}$
is as follows. For the optimal extension for
$E_{sq}^{c}$, $\rho_{ABE}=\sum p_i\rho_{AB}^{i}\otimes
(|i\>\<i|)_{E}$, we have a four-partite state $
\rho_{AA'BB'}=\sum p_i\rho_{AB}^{i}\otimes
(|i\>\<i|)_{A'}\otimes(|i\>\<i|)_{B'}, \nonumber $ then
$I(AA':BB')-I(A':B')=\sum_ip_iI(\rho_{AB}^{i})$.

Once we have the order of the above three measures, we
can easily demonstrate that $C_I$ is lockable i.e. that
one can decrease it about  arbitrary value while removing
a single qubit \cite{DiVincenzo,Karol,CW}. The example is
the {\it flower state} $\rho_{A_1A_2B_1B_2}$
\cite{Karol,CW} defined by its purification:
 \be
|\Psi\>^{A_1A_2B_1B_2C}={1\over
\sqrt{2d}}\sum_{{i=1\ldots
d}\atop{j=0,1}}|i\>^{A_1}|j\>^{A_2}|i\>^{B_1}|j\>^{B_2}U_j|i\>^{C},
\nonumber \ee where $U_0=I$ and $U_1$ is the Fourier
transformation of the computational basis $\{|i\>\}$. It
is shown in \cite{CW} that $E_{sq}^{q}=1+{1\over 2}\log
d$ and furthermore the optimal extension is trivial (the
state itself) that is also one of extensions for $C_I$
and $E_{sq}^{c}$. If $A_2$ is lost, then
$\rho_{A1:B_1B_2}$ is separable. From {\it Prop
\ref{se:order}}, we immediately obtain that $C_I$ and
$E_{sq}^{c}$ are lockable.

It should be emphasized that one is unable to prove
$E_{sq}^{c}$ to be additive at present, but the three
measures are so similar that they are probably the same.
If it is the case, we would have a really graceful result
that the optimal extension is always the classical one.
Moreover, it would give us a strong hint for the
additivity of entanglement of formation that relates to
many other important problems \cite{Shor}. Presumably
$C_I$ may play a role as a bridge.

Among existing bipartite entanglement measures
\cite{4H,PV}, only the relative entropy of entanglement
and the squashed entanglement can be extended to
multipartite cases. Attractively, there exist two
versions of multipartite quantum mutual information
\cite{Cerf}. All conclusions for the bipartite case can
be similarly deduced.

We then obtain two multipartite versions of $C_I$: \bea
C_{I}=\inf\{I_n(A_1A'_1:\cdots:A_nA'_n)-I_n(A'_1:\cdots:A'_n)\},\nonumber\\
C_{S}=\inf\{S_n(A_1A'_1:\cdots:A_nA'_n)-S_n(A'_1:\cdots:A'_n)\},\nonumber
\eea where two candidates for multipartite mutual
information are defined as $ I_n=\sum_iS(A_i)-S(A_1\cdots
A_n)$, and $S_n=\sum_iS(A_1\cdots A_{i-1}A_{i+1}\cdots
A_n)-(n-1)S(A_1\cdots A_n).$ {\proposition ~} The
conditional entanglement for multipartite mutual
information is additive. \bea C_{I}(\rho_{A_1\cdots
A_n}\otimes\sigma_{B_1\cdots B_n})
=C_{I}(\rho_{A_1\cdots A_n})+C_{I}(\sigma_{B_1\cdots B_n}),\nonumber\\
C_{S}(\rho_{A_1\cdots A_n}\otimes\sigma_{B_1\cdots B_n})
=C_{S}(\rho_{A_1\cdots A_n})+C_{S}(\sigma_{B_1\cdots
B_n}).\nonumber \eea

In summary, we have developed a generic approach to
construct new entanglement measures based on quantum
conditioning.  The new measures can not only be obtained
from the known measures but also be generated from
measures of correlations. In particular, a new additive
measure is constructed and generalized to multipartite
entanglement. Moreover, the known additive
measure---squashed entanglement is shown to come from the
asymmetric conditioning. We conjecture that the measures
built from quantum conditioning are additive, which means
that quantum conditioning leads to additive entanglement.
Conditional entanglement measures from other candidates
and further properties will be addressed elsewhere.

{\it Acknowledgement} D. Yang and Z. D. Wang acknowledge
the support by the RGC of Hong Kong (HKU 7051/06 and HKU
3/05C), the URC fund of Hong Kong, the National Natural
Science Foundation of China (10429401), and the State Key
Program for Basic Research of China (2006CB0L1001). M.
Horodecki acknowledge funding by  the Polish Ministry of
Science and Information Technology - grant
PBZ-MIN-008/P03/2003 and the EU IP SCALA.

%---------------------------------------------------------------------------------------

\appendix
\section{Appendix}

{\it Proof of the asymptotic continuity of $C_I$}

The proof is similar to the continuity of
the squashed entanglement \cite{Christandl} that is based
on a basic result in \cite{AF} asserting that for any two
states $\rho_{AB}$ and $\sigma_{AB}$ on ${\cal
H}_{A}\otimes {\cal H}_{B}$, if
$|\rho_{AB}-\sigma_{AB}|=\epsilon$, then \be
|S(A|B)_{\rho}-S(A|B)_{\sigma}|\le4\epsilon\log{d_A}+2H(\epsilon)\label{se:cont},
\ee where $d_A$ is the dimension of ${\cal H}_{A}$ and
$H(\epsilon)=-\epsilon\log{\epsilon}-(1-\epsilon)\log{(1-\epsilon)}$.
Note that the righthand of Eq (\ref{se:cont}) does not
explicitly depend on the dimension of ${\cal H}_{B}$.
Iteratively using the relations between fidelity and
trace norm \cite{Fuchs}, if $|\rho_{AB}-\sigma_{AB}|\le
\epsilon$, then the fidelity $F(\rho_{AB},\sigma_{AB})\ge
1-\epsilon$, then there exist purifications $\Phi_{ABC}$
and $\Psi_{ABC}$ of $\rho_{AB}$ and $\sigma_{AB}$
respectively such that $F(\Phi_{ABC},\Psi_{ABC})\ge
1-\epsilon$, and then $|\Phi_{ABC}-\Psi_{ABC}|\le
2\sqrt{\epsilon}$. For any quantum operation ${\cal E}$
acting on $C$ into $A'B'$, it creates the extensions
$\rho_{AA'BB'}$ and $\sigma_{AA'BB'}$ of $\rho_{AB}$ and
$\sigma_{AB}$ satisfying
$|\rho_{AA'BB'}-\sigma_{AA'BB'}|\le 2\sqrt{\epsilon}$.
Notice that
$I(AA':BB')-I(A':B')=S(A|A')+S(B|B')-S(AB|A'B')$, we get
\bea
&~&|[I(AA':BB')_{\rho}-I(A':B')_{\rho}]\nonumber\\
&-&[I(AA':BB')_{\sigma}-I(A':B')_{\sigma}]|\nonumber\\
&=&|[S(A|A')_{\rho}-S(A|A')_{\sigma}]+[S(B|B')_{\rho}-S(B|B')_{\sigma}]\nonumber\\
&-&[S(AB|A'B')_{\rho}-S(AB|A'B')_{\sigma}]\nonumber\\
&\le&
|S(A|A')_{\rho}-S(A|A')_{\sigma}+|S(B|B')_{\rho}-S(B|B')_{\sigma}|\nonumber\\
&+&|S(AB|A'B')_{\rho}-S(AB|A'B')_{\sigma}|\nonumber\\
&\le&
16\sqrt{\epsilon}\log{(d_Ad_B)}+6H(2\sqrt{\epsilon})=\epsilon'
\eea For a sequence of operation ${\cal E}_i$ that
creates a sequence of extensions such that
$I(AA':BB')_{\rho}-I(A':B')_{\rho}\to E_{I}(\rho_{AB})$,
we have
$|C_I(\rho_{AB})-[I(AA':BB')_{\sigma}-I(A':B')_{\sigma}]|\le\epsilon'$,
then $C_I(\sigma_{AB})\le
I(AA':BB')_{\sigma}-I(A':B')_{\sigma}\le
C_I(\rho_{AB})+\epsilon'$. Similarly $C_I(\rho_{AB})\le
C_I(\sigma_{AB})+\epsilon'$, so $|C_I(\rho_{AB})-
C_I(\sigma_{AB})|\le\epsilon'$.

Notice that we have $\sqrt{\epsilon}$ instead of
$\epsilon$, but it does not change the essence of
condition referring asymptotic continuity \cite{SH}.

{\it Definition of $E_{sq}^{c}$} \cite{MH} The c-squashed
entanglement $E_{sq}^{c}$ is defined as \be
E_{sq}^{c}(\rho_{AB})=\inf{1\over 2} I(A:B|E) \ee where
infimum is taken over the extension states of the form
$\sum p_i\rho_{AB}^{i}\otimes (|i\>\<i|)_{E}$.

In deed, it is equivalent to the mixed convex roof of the
quantum mutual information, i.e. \be
E_{sq}^{c}(\rho_{AB})=\min{1\over 2}\sum_ip_iI(\rho_{AB}^{i}), \ee
where $\rho_{AB}=\sum_ip_i\rho_{AB}^{i}$.

\end{document}